\documentclass[11pt,twoside]{article}

\usepackage{asp2014}

\aspSuppressVolSlug
\resetcounters

\begin{document}

\title{Modernizing IRAF to Support Gemini Data Reduction}

\author{Michael Fitzpatrick,$^1$ Vinicius Placco,$^1$ Adam Bolton,$^1$ Brian
Merino,$^1$ Susan Ridgway,$^1$ and Letizia Stanghellini$^1$}
\affil{$^1$NSF's NOIRLab, Tucson, AZ, USA; \email{mike.fitzpatrick@noirlab.edu}}

\paperauthor{Michael Fitzpatrick}{mike.fitzpatrick@noirlab.edu}{0000-0002-9080-0751}{NSF's NOIRLab}{}{Tucson}{AZ}{85719}{USA}
\paperauthor{Vinicius Placco}{vinicius.placco@noirlab.edu}{0000-0003-4479-1265}{NSF's NOIRLab}{}{Tucson}{AZ}{85719}{USA}
\paperauthor{Adam Bolton}{adam.bolton@noirlab.edu}{}{NSF's NOIRLab}{}{Tucson}{AZ}{85719}{USA}
\paperauthor{Brian Merino}{brian.merino@noirlab.edu}{}{NSF's NOIRLab}{}{Tucson}{AZ}{85719}{USA}
\paperauthor{Susan Ridgway}{susan.ridgway@noirlab.edu}{}{NSF's NOIRLab}{}{Tucson}{AZ}{85719}{USA}
\paperauthor{Letizia Stanghellini}{letizia.stanghellini@noirlab.edu}{}{NSF's NOIRLab}{}{Tucson}{AZ}{85719}{USA}



\begin{abstract}
 The US National Gemini Office (US NGO), part of the Community Science and Data Center (CSDC) at NSF’s NOIRLab, has completed a project to upgrade the IRAF-based Gemini reduction software to provide a fully supported system capable of running natively on modern hardware. This work includes 64-bit platform ports of the GEMINI package and dependency tasks (e.g. from the STSDAS external package), upgrades to the core IRAF system and all other external packages to fix any platform and licensing problems, and the establishment of fully supported Help Desk and distribution systems for the user community.

	 Early results show a 10-20X speedup of execution times using the native 64-bit software compared to the virtualized 32-bit solutions now in use.  Results are even better on new Apple M1/M2 platforms where the additional overhead of Intel CPU emulation can be eliminated. Timing comparisons, science verification testing, and release plans are discussed.
\end{abstract}



\section{Introduction}
	Using legacy software to support ongoing observatory operations only becomes more difficult with time. Hardware and software environments evolve, adding to maintenance costs and further limiting the options available to run legacy code,and potentially leading to lapses in support for key observing modes while replacement systems are being developed.  Such is the case for Gemini data reduction today, especially in the case of spectroscopic instruments awaiting new pipelines in the DRAGONS \citep{2019ASPC..523..321L} software system.  
	
	While the IRAF  \citep{10.1117/12.968154}  system has been 64-bit compatible for a number of years \citep{fitzpatrick2012iraf}, it quickly fell behind the requirements of modern operating systems due to a lack of ongoing maintenance.  The GEMINI external package remained 32-bit-only and so users relied on continued vendor support of 32-bit platforms or virtualization solutions for those instruments and modes not yet supported by DRAGONS. For the Gemini user community reliant on these tools the combination of these issues would only worsen over time.
	
	To address this problem, CSDC's US NGO initiated an IRAF  modernization project with a well-defined and limited scope:
\begin{itemize}
    \item Port the GEMINI reduction package and its dependencies to 64-bit,
    \item Fix all OS, platform, and licensing issues in existing code,
    \item Provide ongoing User Support and distribution until suitable replacements for all Gemini instruments and modes are available.
 \end{itemize}

No new development of the core system is planned beyond work required to meet the above goals. Updates to the PyRAF-based reductions were explicitly \emph{not} part of these modernization efforts.

\section{Implementation Plan}
The project implementation was staged to maximize time available for testing and code/science verification:
\begin{itemize}
\item Identify and extract GEMINI package dependency code to a new support package
\item Port GEMINI package and dependency code to 64-bit
\item Update core system and external packages with needed OS and platform fixes
\item Remove or replace any code with license issues
\end{itemize}
Once the first two stages were complete, science testing of the ported code could begin using existing 64-bit IRAF systems to find those tasks which no longer worked and to verify the results produced were the same as those from the existing 32-bit GEMINI tasks.  This work was done in parallel with the remaining stages and revisited at each later stage. Updating the core IRAF system was done independently of external package testing and represented the bulk of the effort on this project.

\section{Performance Improvements}
	As expected, a native 64-bit port of the GEMINI tasks and optimization of the core system has lead to significant reduction in the time taken to reduce data as compared to the existing virtualized approach (hereafter \emph{GemVM} \citep{GemVM} ).  Figure 1 shows the times required to execute the test tasks on various platforms and Figure 2 shows the relative speedup of those tasks.  The variability in relative time can be explained by noting that some reductions are more I/O intensive and may thus strain the older hardware used.
	
	The GemVM reference platform was run on the fastest machine available (a 2023 MacBook Pro M2/Max) and had the additional overhead of CPU emulation, however this still realistically reflects how the platform is used in the community.  Faster speeds have been reported with the GemVM running on native Intel hardware which we attribute to the elimination of the Intel emulation by the VM on those platforms.  Lastly, PyRAF versions of some scripts were similarly tested and show 10-20\% slower times due to the additional emulation of the IRAF Command Language (CL) required.

\articlefigure{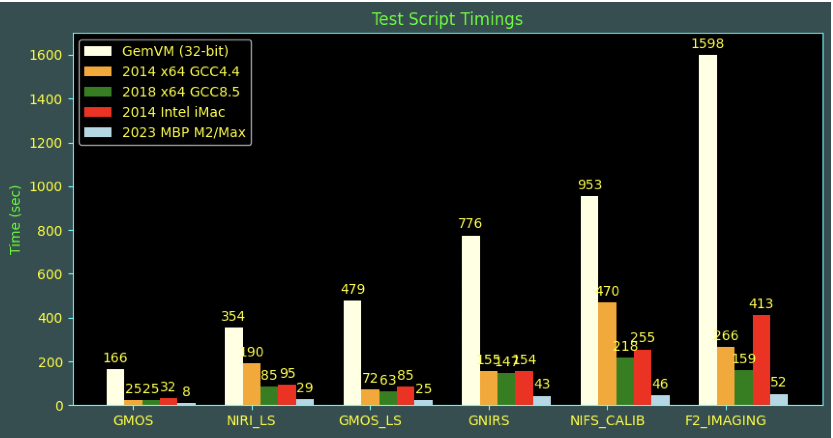}{ex_fig1}{Comparison of runtimes for various reduction test scripts on  multiple machines showing  improved runtimes using the new native IRAF v2.18 system as compared to the Gemini VM.}
\articlefigure{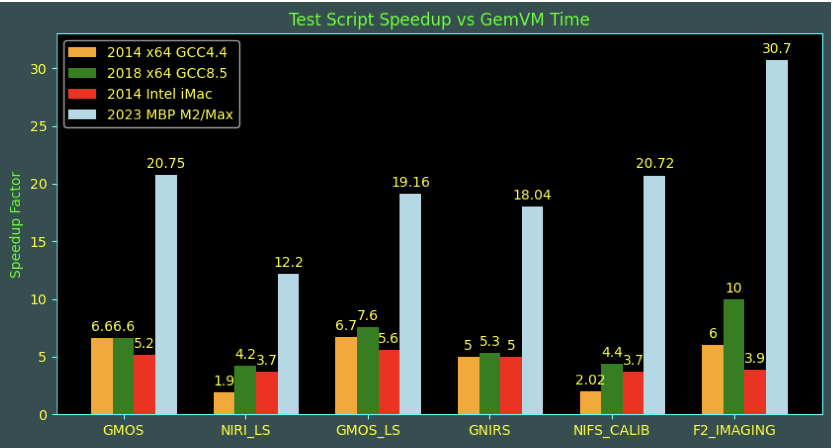}{ex_fig2}{Relative speedup of runtimes as compared to the Gemini VM.}

\section{Testing and Code Verification}
	The GEMINI package conveniently contains example reduction scripts that utilize public Gemini  archive data as part of the documentation of each instrument package.  These scripts were used to test the new code at each stage of implementation as a means to exercise multiple tasks in the package at once.  For code testing it was less important that the outputs be the best scientific data product than that they be the same as that produced by the reference platform and that tasks ran without error.  This focus on reproducibility versus correctness allowed testers not familiar with the details of Gemini data reduction to still exercise the tasks to find problems. Specific scripts for the STSDAS dependency tasks were not available, however those tasks were used indirectly within the example scripts and thus similarly tested.  
	
		This testing allowed us to fix a number of bugs in the existing core IRAF code as well as errors within the test script and bugs introduced by the IRAF and GEMINI update itself. Science verification will be ongoing as the new system is used to carefully reduce and analyze data by community users.
		
	The core IRAF system lacks a comprehensive test package and so targeted testing was used on specific code changes where needed and overall system functionality (e.g. WCS handling) was examined through spot checking the most commonly used  tasks that demonstrate the function being tested.
	
\begin{table}[!ht]
\caption{Test scripts used in timing tests.}
\begin{center}
{\small
\begin{tabular}{rll}  
\tableline
\noalign{\smallskip}
Name & Description & Size\\
\noalign{\smallskip}
\tableline
\noalign{\smallskip}
GMOS &  Basic reduction/extraction of GMOS science frames & 4 images, 17MB \\
GMOS\_LS & Typical reduction of standard-star observation & 8 images, 28MB \\
GNIRS & GNIRS basic reductions & 53 images, 213MB \\
NIRI\_LS & Typical reduction of NIRI spectroscopy data & 43 images, 183MB \\
NIFS\_CALIB & Baseline NIFS calibration & 10 images, 164MB \\
F2\_IMAGING & Typical reductions for F2 imaging and calib data & 57 images, 938MB \\
\noalign{\smallskip}
\tableline
\end{tabular}
}
\end{center}
\end{table}

\section{Simplified Installation Procedures}
This project also took the opportunity to simplify the installation process for all users by recognizing that the most common user today would be a scientist working on a single-user laptop or desktop system.  This allowed us to create drag-and-drop installers for Mac systems and self-extracting installation scripts for Linux to ensure access for users of all skill levels.  
Downloads are available from the project website at \texttt{https://iraf.noirlab.edu}.

\acknowledgements  This work is supported by NOIRLab, which is managed by the Association of Universities for Research in Astronomy (AURA) under a cooperative agreement with the National Science Foundation. In particular, US NGO staff were critical in science testing the new GEMINI package to ensure proper results. We are grateful to our colleagues at Gemini for useful discussions and internal access to their package test environment.
	
	This project also benefited greatly from the earlier work done by the ‘iraf-community’ \citep{iraf-community} project to identify and fix bugs in the core IRAF system.  

\bibliography{P914}  


\end{document}